\documentclass{article}[12pts]
\title{A note on the Lee-Yang circle theorem}
\author{Ranjan Kumar Ghosh\thanks{\normalsize e-mail address:- rkg\_1978@yahoo.com}\\Bidhannagar College\\EB-2, Sector-1, Salt Lake City\\Kolkata-700 064, INDIA}
\date{December22, 2011}
\begin{document}
\maketitle
\begin{abstract}\normalsize
A simple proof of the celebrated theorem of Lee and Yang is attempted in this short note
\end{abstract}
\large
The celebrated theorem due to Lee and Yang $\left[1\right]$states that all the zeroes of the grand partition of the two-dimensional lattice gas or the partition function of the ferromagnetic Ising model in a magnetic field lie on the unit circle when written in terms of an appropriate variable.It is one of the very few rigorous exact results that are available in statistical mechanics. There have been many generalizations of this result in the context of different statistical mechanical models$\left[2\right]$.  The proof of the result is however rather complicated and involved. In this short note I try to give a simple and rather transparent proof of the result obtained by Lee and Yang.\\
 Before that we give the statement as given in their paper.\\
\textit{Theorem}: Let $x_{\alpha\beta}=x_{\beta\alpha} \left(\alpha\neq\beta,\alpha,\beta=1,2,\dots,n\right)$ be real numbers \textit{whose absolute values are less than or equal to 1}. Divide the integers 1,2,$\dots$,n into two groups a and b so that there are $\gamma$ integers in group a and $\left(n-\gamma\right)$ in group b.Consider the product of all $x_{\alpha \beta}$ where $\alpha$ belongs to group a and $\beta$ belongs to group b. We shall denote by $P_{\gamma}$ the sum of all such products over all the $n!\over{\gamma!\left(n-\gamma\right)}!$ possible ways of dividing the $n$ integres. In other words 
\begin{equation}P_{\gamma}=\sum\left[\gamma!\left(n-\gamma\right)!\right]^{-1}\prod_{j=1}^{n-\gamma}\prod_{i=1}^{\gamma}x_{a_{i}b_{j}}\end{equation}
where$a_{1},\dots,a_{\gamma},b_{1}\dots b_{\left(n-\gamma\right)}$ is any permutation of the integers1,2,$\dots$,n and summation extends over all such permutations:e.g.
\begin{equation}P_{0}=P_{n},P_{1}=P_{n-1}=x_{12} x_{13}\dots x_{1n}+x_{21} x_{23}\dots x_{2n}+\dots +x_{n1} x_{n2}\dots x_{n\left(n-1\right)}\nonumber \end{equation}
It is easy to verify  that $P_{\alpha}=P_{n-\alpha}$.
Consider the polynomial

\begin{equation}{\cal P} \left(z\right)=1+P_{1}z+P_{2}z^{2}+\dots+P_{n-1}z^{n-1}+z^{n}\end{equation}
The theorem asserts that all the roots of the equation
 \begin{equation}{\cal P}= 0 \nonumber\end{equation}
lie on the unit circle.
In the proof of the above result it is assumed that none of the $z_{n}$'s has an absolute value less than one.\\
As stated by Griffiths $\left[3\right]$ the theorem of Lee and Yang for the Ising ferromagnets has the following form:(for a complete statement, please  look up the article)\\
\textit {Theorem:  (a) Provided all the interactions $J_{ij}$ satisfy
\begin{equation}J_{ij}>0\end{equation}
and provided\begin{equation}\left|z_{i}\right| \geq 1 \end{equation} for all $i,j$ then $Z=0 $ implies that \begin{equation} \left|z_{1}\right|=\left|z_{2}\right|= \dots= \left|z_{\nu} \right|=1 \end{equation}
(b) Provided the interactions $J_{ij}$ satisfy
\begin{equation}J_{ij}\geq 0\end{equation}
and provided all the $z_{i}$ are equal to $z$, $Z=0$ iomplies $\left|z\right|=1$}\\
For the proof Griffiths refers to the original paper by Lee and Yang which is ref.[1]. It is the part (b) of this statement whose proof we seek in this note.\\
The equation(4) written explicitly is
\begin{equation}1+P_{1}z\dots+P_{n-1}z^{n-1}+ z^{n}=0\end{equation}
If $ z_{1}$  is a zero of the polynomial then we must have
\begin{equation}1+P_{1}z_{1}+\dots+P_{n-1}z_{1}^{n-1}+z_{1}^{n}=0.\end{equation}
Taking the complex conjugate of above expression we get
\begin{equation} 1+P_{1}^{*}z_{1}^{*}+\dots+P_{n-1}^{*}z_{1}^{*\left(n-1\right)}+z_{1}^{*n}=0\end{equation}
Which shows that $z_{1}^{*}$ is also a zero of the same expression if the coefficients of the equation are real. This is the well known result that for an equation with real coefficients the complex roots occur in conjugate pairs.\\
Now let us assume that the given polynomial has its zeroes at $z_{1},z_{2},\dots,z_{n}$ and hence can be written in the factorized form \begin{equation}\left(z-z_{1}\right)\left(z-z_{2}\right)\dots\left(z-z_{n}\right)\end{equation}
Expanding this expression and comparing it with the original polynomial we see that 
\begin{equation}\left(-1\right)^{n} z_{1} z_{2}\dots z_{n}=1\end{equation}
Taking the modulus of this expression we have
\begin{equation}\left|z_{1}\right|\left|z_{2}\right|\dots\left|z_{n}\right|=1\end{equation} 
Now this expression shows that if $\left|z_{k}\right|$ for any $k$ is larger than one then to satisfy this equation, at least one or more of $\left|z_{m}\right|$ for any $m$ different from $k$ must be less than one. This contradicts the assumption in the beginning of this note that none of $\left|z_{n}\right|$ is smaller than one. Hence all the $z_{n}$'s must have absolute value unity or in other words each $z_{n}$ must lie on the unit circle in the complex $z$ plane. This completes the proof of the assertion.\\
It should be noted that in arriving at eq.(13) we have nowhere assumed any specific structure for the coefficients of other terms, not even the assumption that they must be real. Hence the result is correct even in the case when the other coefficients are complex, as long as the heighest degree term of the polynomial is $z^{n}$ and the constant term is equal to unity (or a pure phase $e^{i\alpha}$). This can be taken as a generalization of the Lee-Yang result.\\
It is also seen that upon imposing the reality conditions on the coefficients of the polynomial, we must have the roots in complex conjugate pairs, or in other words there always are two roots which are equidistant from the real axis on either side of it. Hence as one of them moves towards the real axis from the upper half of the complex plane the other one must do the same from the other side ,i.e. the lower half of the complex plane.\\
 To summarise, we have given a rather simple proof of the original Lee-Yang circle theorem that may be important in understanding the other results related to it.

\end{document}